\documentclass[conference]{IEEEtran}
\IEEEoverridecommandlockouts
\usepackage{cite}
\usepackage{amsmath,amssymb,amsfonts}
\usepackage{graphicx}
\usepackage{textcomp}
\usepackage{xcolor}
\usepackage{comment}
\usepackage{algorithm}
\usepackage{algpseudocode}
\usepackage{multirow}

\usepackage{url}
\usepackage{varwidth}
\usepackage{microtype}

\usepackage{hyperref}

\def\BibTeX{{\rm B\kern-.05em{\sc i\kern-.025em b}\kern-.08em
    T\kern-.1667em\lower.7ex\hbox{E}\kern-.125emX}}

\graphicspath{{./images/}} 

\begin{document}
\title{{\sc Zeus}: An Efficient GPU Optimization Method Integrating PSO, BFGS, and Automatic Differentiation}

\author{
\IEEEauthorblockN{%
Dominik So\'os\IEEEauthorrefmark{1},
Marc Paterno\IEEEauthorrefmark{1}\IEEEauthorrefmark{2}\thanks{Notice: This work was produced by FermiForward Discovery Group, LLC under Contract No. 89243024CSC000002 with the U.S. Department of Energy, Office of Science, Office of High Energy Physics. The United States Government retains and the publisher, by accepting the work for publication, acknowledges that the United States Government retains a non-exclusive, paid-up, irrevocable, world-wide license to publish or reproduce the published form of this work, or allow others to do so, for United States Government purposes. The Department of Energy will provide public access to these results of federally sponsored research in accordance with the DOE Public Access Plan (http://energy.gov/downloads/doe-public-access-plan). FERMILAB-PUB-25-0142-CSAID},
Desh Ranjan\IEEEauthorrefmark{1},
Mohammad Zubair\IEEEauthorrefmark{1}}
\IEEEauthorblockA{\IEEEauthorrefmark{1}Department of Computer Science \\
Old Dominion University \\
Norfolk, Virginia, USA}
\IEEEauthorblockA{\IEEEauthorrefmark{2}Fermi National Accelerator Laboratory \\
Batavia, Illinois, USA \\ }
}

\maketitle

\begin{abstract}
We introduce a novel, efficient computational method, {\sc \textbf{Zeus}}, for numerical optimization, and provide an open-source implementation.
It has four key ingredients: (1) particle swarm optimization (PSO), (2) the use of the Broyden-Fletcher-Goldfarb-Shanno (BFGS) method, (3) automatic differentiation (AD), and (4) GPUs.
Our approach addresses the computational challenges inherent in high-dimensional, non-convex optimization problems.
In the first phase of the algorithm, we get a potentially good set of starting points using PSO. 
Thereafter, we run BFGS independently in parallel from these starting points. 
BFGS is one of the best-performing algorithms for numerical optimization. However, it requires the gradient of the function being optimized.
{\sc \textbf{Zeus}} integrates automatic differentiation into BFGS thus avoiding the need for the user to calculate derivatives explicitly.
The use of GPUs allows {\sc \textbf{Zeus}} to speed up the calculations substantially.
We carry out systematic studies to explore the trade-offs between the number of PSO iterations taken, starting points, and BFGS iteration depth.
We show that a handful of iterations of PSO can improve global convergence when combined with BFGS. 
We also present performance studies using common test functions.
The source code can be found at \href{https://github.com/fnal-numerics/global-optimizer-gpu}{https://github.com/fnal-numerics/global-optimizer-gpu}.

\end{abstract}


\section{Introduction}
A wide range of domains make use of numerical optimization, including particle physics simulations, machine learning, and financial modeling.
Multidimensional non-convex global optimization can be difficult due to a number of reasons.
One of these reasons is that the solution space grows exponentially with the dimensionality of the function being optimized.

The classical sequential numerical optimization algorithms start from an initial ``guess" and then use the function gradient to move in the direction that reduces the value of the objective function. 
However, such conventional optimization techniques often fail to navigate complex landscapes with many local minima or narrow valleys where the gradient is close to zero.
Purely gradient-based methods like stochastic gradient descent \cite{ketkar2017stochastic}, mini-batching \cite{li2014efficient, singh2024mini}, and stochastic variance-reduced gradient \cite{reddi2016stochastic} get stuck in flat regions or local minima when dealing with non-convex landscapes. 

To tackle difficult landscapes with many local minima, the particle swarm optimization (PSO) method has been widely adopted for its ability to handle global searches by exploring multiple regions of the hyperspace.
However, PSO by itself struggles with problems with flat regions.

The use of graphical processing units (GPUs) has become popular in parallel computing, as they can run thousands of computations simultaneously. Parallel computing approaches have been proposed that distribute the computations of the gradients \cite{zinkevich2010parallelized}. 

This paper investigates how parallel computing on GPUs can improve non-convex global optimization strategies.
We propose a novel, GPU-based global optimization algorithm named {\sc Zeus} that combines the particle swarm optimization (PSO) with popular quasi-Newton optimization (BFGS) \cite{broyden1970convergence, fletcher1970new, goldfarb1970family, shanno1970conditioning}. Additionally, the method provides a built-in forward-mode automatic differentiation (AD) library, thus avoiding the explicit calculation of gradients needed in the BFGS method.
We propose an approach that initializes the optimization from a multitude of random starting points in multidimensional solution space, utilizing GPUs to concurrently execute the BFGS optimization algorithm using forward-mode AD. 
Our contributions are summarized below.
\begin{itemize}
    \item We provide a parallel multistart optimization algorithm with automatic differentiation that combines the advantages of the PSO and BFGS methods as well as automatic computation of AD for accuracy and ease of use. 
    \item Our C++ CUDA implementation of the algorithm provides a 10- to 100-fold  speedup compared to our serial implementation.
    \item We provide insight into the trade-off between the number of PSO steps taken and the number of optimizations run concurrently through our extensive studies. 
\end{itemize}

Our extensive experiments not only confirm that a handful of PSO iterations can boost global convergence rates, as shown in Figure~\ref{fig:pso_iter}, but also reveal hyperparameter trade-offs for different functions.
We further discuss limitations like the failure condition on functions like the Ackley function with discontinuous gradients.
We outline future plans to reduce computational complexity in the parts of the algorithm that dominate runtime.


The rest of the paper is organized as follows: In the next section, we present some background and previous related work.
In Section~\ref{sec:three}, we present the sequential {\sc Zeus} algorithm.
Section~\ref{sec:four} presents the parallel version of the algorithm.
Section~\ref{sec:five} presents our experiments and results.
The last section discusses our results and observations as well as future work.
In the next section, we present the background information and the related work to our approach, where we discuss each component of the algorithm, including multistart and swarm-based algorithms, BFGS with AD, and GPUs.

\section{Background and Related work}
Stochastic multistart and evolutionary algorithms, where the optimizer starts from many points, are not a new concept. 
The earliest strategies date back to the 1970s, the first multistart method, which proposed the idea of launching many optimizations from many different points to improve the chance of finding the global optimum \cite{goldstein1971descent,boender1982stochastic} and was later implemented as GLOBAL \cite{csendes1988nonlinear}.

The PSO method is widely used in the literature to improve global convergence.
\cite{li2011hybrid} combined PSO and BFGS, but run on a CPU and do not leverage automatic differentiation or parallel GPUs.
Other approaches used a two-phase global-local scheme, but neither used PSO nor BFGS, nor AD \cite{ferreiro2019parallel}.
Similarly, \cite{barkalov2016parallel} implemented a GPU-based global search by space-filling, but no PSO, BFGS, or AD was used. 
Our unique combination is the first C++ CUDA-based library to integrate:
\begin{itemize}
    \item global search with PSO
    \item local refinement via BFGS
    \item automatic differentiation for gradient calculation
    \item massively parallel GPU
\end{itemize}

The well-established BFGS algorithm that originates from the works of Broyden \cite{broyden1970convergence}, Fletcher \cite{fletcher1970new}, Goldfarb \cite{goldfarb1970family}, and Shanno \cite{shanno1970conditioning} uses quasi-Newton updates to approximate the Hessian and typically converges in fewer iterations than first-order methods. 
\cite{pu1990convergence} showed that BFGS combined with a Wolfe line search enjoys global convergence properties even in moderately high dimensions, further motivating its application to non-convex problems.

A major challenge with gradient-based methods, however, is the need for accurate derivative information. Manual derivation is error-prone and often impractical for complex, high-dimensional functions. This is where AD becomes invaluable.

\subsection{Automatic Differentiation}
In recent years, AD has gained widespread attention, not only in machine learning but also in scientific computing \cite{hueckelheim2023short,baydin2018automatic,bucker2006automatic,margossian2019review}.
AD is attractive because it provides more accurate derivatives with minimal overhead without the need for the user to provide manual derivation of the gradient of the objective function.
For instance, \cite{zubair2023efficient} implemented a forward-mode AD technique on GPUs for computational fluid dynamics, achieving performance close to the hardware’s peak throughput with higher accuracy than manual derivative calculation.
Likewise, \cite{grabner2008automatic} used AD for 2D/3D registration on GPUs, addressing challenges in medical imaging.

Recent advances have further accelerated AD on GPUs using modern compiler techniques. 
Google Brain's JAX framework \cite{jax2018github} offers a user-friendly, NumPy-like interface that supports just-in-time, or JIT, compilation and GPU acceleration while providing robust automatic differentiation. 
However, their implementation is not accessible within user-defined GPU kernels.
Despite these advances, most available AD libraries, such as those in DLib by \cite{king2009dlib} or the Stan Math Library \cite{carpenter2015stanmathlibraryreversemode}, still predominantly target CPU execution.

\subsection{Swarm-Intelligence Algorithms}
Another type of multistart algorithm related to our work is Particle Swarm Optimization (PSO), first proposed by Kennedy and Eberhart \cite{kennedy1995particle} then later improved by Shi and Eberhart \cite{shi1998modified} and further stabilized by  Clerc and Kennedy \cite{clerc2002particle}.
It was introduced as a population‐based stochastic optimization technique inspired by social behaviors in animals.
PSO algorithms use many particles that move toward promising regions in the search space based on information shared between the particles \cite{kennedy1995particle}.

Different variants of the PSO algorithm have been developed \cite{jain2022overview} that utilize AD \cite{della2024automatic,noel2012new,thobirin2015automatic}, or BFGS \cite{li2011hybrid,wu2014superior,zhang2016hybrid,nezhad2013particle}, or LBFGS using GPUs \cite{dixit2024efficient}, but not all four together.
While the PSO approach attempts to converge to a global minimum, each particle is influenced by the others.
It does not guarantee global convergence. 
Instead of each optimization moving to a single point as a swarm, we explore the search space while also adjusting the velocity of each particle based on the global best. 
Although multistart and swarm-based techniques have been around for a long time, GPU-accelerated AD and quasi-Newton methods like BFGS have not been used together much. 
Our approach is designed to handle complex non-convex landscapes like Rastrigin and Rosenbrock functions by running many separate optimizations at the same time from a variety of random starting points. 
We only consider GPU-accelerated algorithms for comparison.


\subsection{GPU-Accelerator}
ParallelParticleSwarms \cite{dixit2024efficient} is the most related work to our work since they combine GPU with BFGS and SciML allows for AD computation. 
However, after extensive communication with the authors, we were unable to use their hybrid method.

While multistart and swarm-based algorithms have been extensively studied, and recent advancements have accelerated automatic differentiation on GPUs using modern compiler techniques such as Enzyme \cite{moses2021reverse}, a comprehensive integration of PSO, forward-mode AD with the BFGS algorithm on GPUs remains under-explored. 
Our work addresses this gap by concurrently executing multiple GPU-accelerated BFGS optimization threads initialized by PSO, leveraging forward-mode AD.
This approach not only enhances computational efficiency, but also improves the likelihood of converging to the global minimum. 
In the next section, we provide an overview of the details of the sequential {\sc Zeus} algorithm. 



\section{{\sc Sequential Zeus} Algorithm} \label{sec:three}
The main methodology implements a hybrid PSO with a gradient-based BFGS using AD for the gradient calculation at each iteration. 
Algorithm~\ref{alg:sequentialzeus} describes the sequential version. 
The bold variables indicate that they are vectors of size $N\times dim$ or $dim$, where $N$ is the number of particles, and $dim$ is the dimension of the objective function.
The algorithm can be broken into two main phases.

In the first phase, using Algorithm~\ref{alg:initpso} we initialize the $\mathbf{swarm}$ of size $N$, where $\mathbf{swarm}$ stores the current coordinates for each particle for each dimension.
We assign $\mathbf{V}$ to each particle to hold the velocities in each dimension.
Then, using Algorithm~\ref{alg:updateswarm} we update the swarm using each particle's best position $\mathbf{pX}$ with its best value $pB$ and the global best $gB$ for each iteration.

The second phase is Algorithm~\ref{alg:sequential-bfgs}, which is the sequential BFGS.
We introduce a convergence criterion $\mathit{required_c}$, which holds the number of required convergences set by the user. 
We use Algorithm~\ref{alg:forwardad} to calculate the derivatives for each dimension using our automatic differentiation library. 
Then, we utilize a backtracking line search method by \cite{armijo1966minimization} described in Algorithm~\ref{alg:line-search}, to calculate an optimal step size $\alpha$.
The algorithm will keep looping until we have enough optimizations that have converged to the threshold $\Theta$ set by the user.
If an optimization does not reach an area where the norm of the gradient is sufficiently small in fewer than $iter_{bfgs}$ iterations, then we terminate that optimization with failure status.
The following sections provide an overview of each ingredient of our method. 

\begin{algorithm}[ht]
    \begin{algorithmic}[1]
    \caption{Sequential PSO-BFGS}
    \Function{Sequential{\sc Zeus}}{$f, N,\mathrm{range}, \mathrm{iter}_{pso},\allowbreak \mathrm{iter}_{bfgs}, \Theta, required_{c}, w,\,c_1,\,c_2$}
        \State Allocate $\mathbf{swarm}[N*dim], \mathbf{V}[N*dim], \mathbf{pX}[N*dim], \mathbf{pB}[N], \mathbf{gX}[dim]$
        \State $gB \gets +\infty$
        \State $\mathbf{swarm} \gets $\Call{initSwarm}{$f,N,\mathrm{range},\mathbf{swarm},\mathbf{V},\mathbf{pX},$}
        \State $c \gets 0$;  \Comment{number of converged runs}
        \For{\(j \gets 0\) to $\mathrm{iter}_{pso}-1$}
            \State $\mathbf{swarm} \gets $\Call{updateSwarm}{$f,N,\mathrm{range},\mathbf{swarm},\allowbreak \mathbf{V},\mathbf{pX}, pVal,\mathbf{gX},gF,w,\,c_1,\,c_2$}
        \EndFor
        \ForAll{\(i \gets 0,\dots,N-1\)}
            \State $r \gets \Call{SerialBFGS}{f, \mathrm{range},\mathrm{iter}_{bfgs},\mathbf{swarm[i]},\Theta}$
            \If{$r.fval < gF$}
                \State $gBs \gets r$
            \EndIf
            \If{$r.\mathrm{status} = 1$}
                \State $c \gets c+1$
            \If{$c = \mathrm{required}_c$}
                \State \textbf{break}  \Comment{stop early once enough runs have converged}
            \EndIf
        \EndIf
        \EndFor
    \State \Return estimated global minimum $gB$
    \EndFunction
    \label{alg:sequentialzeus}
\end{algorithmic}
\end{algorithm}


\subsection{Specifying the starting points}
Optimization performances vary greatly depending on the initialization of the algorithm.
Therefore, it is beneficial to explore methods other than random number generation to add intelligence to the algorithm.

\subsubsection{Random Number Generation} \label{sec:random}
For the sequential random number generation, we relied on standard libraries random device and their uniform distribution.
However, starting from random points is suboptimal and can be improved using PSO.

\subsubsection{Randomness improved by PSO}
We developed the serial version of this algorithm that the sequential {\sc Zeus} algorithm is using. 
We integrate PSO into the {\sc Zeus} algorithm as an option to initialize the starting points using random numbers and then use swarm intelligence to improve the location of such starting points. 
The initialization of the swarm is being done in Algorithm~\ref{alg:initpso}.
Then, we update the swarm using Algorithm~\ref{alg:updateswarm}.
Hyperparameter optimization could be its own research project.
In this paper we have used these hyperparameters for the PSO:
$w = 0.5, c_1=1.2, c_2=1.5$, where $w$ is the inertia, $c_1$ is the cognitive coefficient, and $c_2$ is the social coefficient.
Previous work has used the same parameters in their implementation \cite{deboucha2020modified}. 
Once the starting points have been specified, we continue to BFGS.

\begin{algorithm}
\caption{Initialization of the swarm}
    \begin{algorithmic}
    \Function{InitSwarm}{$f,N,range,\mathbf{swarm}, \mathbf{V},\mathbf{pX},pVal,\allowbreak \mathbf{gX}, gB$}
        \State $\,\mathit{lower}\gets \mathrm{range.lower},\;\mathit{upper}\gets \mathrm{range.upper}$
        \State $\,\mathit{vel\_range}\gets(\mathit{upper}-\mathit{lower})$
        \For{$i\gets 0$ \textbf{to} $N-1$}
                \State $\mathbf{swarm[i]}\gets \mathrm{UniformSample}(\mathit{lower},\mathit{upper})$
                \State $\mathbf{V[i]}\gets \mathrm{UniformSample}(-\mathit{vel\_range},\mathit{vel\_range})$
                \State $\mathbf{pX[i]}\gets \mathbf{swarm[i]}$
            \State $pVal[i]\gets f\bigl(\mathbf{swarm[i]})$
            \If{$i=0$ \textbf{or} $pVal[i]<gF$}
                \State $gF\gets pVal[i]$
              \State $\mathbf{gX} \gets \mathbf{pX[i]}$
            \EndIf
        \EndFor
        \textbf{return} $swarm$
    \EndFunction
    \end{algorithmic}
\label{alg:initpso}
\end{algorithm}

\begin{algorithm}
\caption{Sequential function to update each particle’s velocity, position, and bests}
\label{alg:updateswarm}
\begin{algorithmic}
    \Function{UpdateSwarm}{$f,N,\mathbf{swarm},\mathbf{V},\mathbf{pX},pVal,\allowbreak \mathbf{gX},gB,w,\,c_1,\,c_2$}
        \For{$i\gets 0$ \textbf{to} $N-1$}
                \State $\mathbf{r_1}\gets \mathrm{UniformSample}(0,1)$
                \State $\mathbf{r_2}\gets \mathrm{UniformSample}(0,1)$
                \State $\mathbf{x}\gets \mathbf{swarm[i]},\mathbf{v}\gets \mathbf{V[i]},\mathbf{p}\gets \mathbf{pX[i]}$
                \State $\mathbf{g}\gets \mathbf{gX}$
                \State $\mathbf{v'} \gets w\,\mathbf{v} + c_1\,\mathbf{r_1}\,(\mathbf{p} - \mathbf{x})\;+\; c_2\,\mathbf{r_2}\,(\mathbf{g} - \mathbf{x})$
                \State $\mathbf{x'} \gets \mathbf{x} + \mathbf{v'}$
                \State $\mathbf{V[i]} \gets \mathbf{v'},\quad \mathbf{swarm[i]}\gets \mathbf{x'}$
            \State $fval\gets f\bigl(\mathbf{swarm[i]})$
            \If{$fval < pVal[i]$}      \Comment{update personal best}
                \State $pVal[i]\gets fval$
                \State $\mathbf{pX[i]}\gets \mathbf{swarm[i]}$
            \EndIf
            \If{$fval < gF$}           \Comment{update global best}
                \State $gF\gets fval$
                \State $\mathbf{gX} \gets \mathbf{swarm[i]}$
            \EndIf
        \EndFor
        \State \Return $swarm$
    \EndFunction
\end{algorithmic}
\end{algorithm}

\subsection{BFGS}

\begin{algorithm}[ht]
\begin{algorithmic}[1]
\caption{BFGS procedure with forward-mode AD}
\label{alg:sequential-bfgs}
\Function{SerialBFGS}{$f, \mathrm{range},\mathrm{iter}_{bfgs},\mathbf{swarm[i]},\Theta$}
    \State \(\mathbf{x} \gets \mathbf{swarm[i]}\), \(k \gets 0\) \Comment{Set initial guess, loop counter}
    \State  \(\mathbf{H} \gets \mathbf{I}\) \Comment{Identity matrix for Hessian}
    \While{\(k < \mathrm{iter}_{bfgs}\)}
        \State \(\nabla f(x) \gets \) \Call{forwardAD}{$f$, $\mathbf{x}$} 
        \If{\(\|\nabla f(x)\| < \Theta\)}
            \State $result.status = 1$;
             \State \textbf{return} $result$ \Comment{Convergence criterion met}
        \EndIf
        \State \(\mathbf{p} \gets -\mathbf{H}\, \nabla f(\mathbf{x})\) \Comment{Calculate search direction}
        \State \(\alpha \gets \Call{lineSearch}{f(\mathbf{x}),\, \mathbf{x}, \mathbf{p}, \mathbf{g},\mathrm{dim},\mathrm{iter}_{ls}}\)
        \State \(\mathbf{x_{\text{new}}} \gets \mathbf{x} + \alpha\, \mathbf{p}\) \Comment{Update current point}
        \State \(\mathbf{\delta x} \gets \mathbf{x_{\text{new}}} - \mathbf{x}\) \Comment{Compute differences}
        \State \(\mathbf{\delta g} \gets \nabla f(\mathbf{x_{\text{new}}}) - \nabla f(\mathbf{x})\)
        \State Hessian update:
        \[\mathbf{H}_{k+1} \gets \left(\mathbf{I} - \frac{\mathbf{\delta x} \mathbf{\delta g^T}}{\mathbf{\delta x^T} \mathbf{\delta g}} \right) \mathbf{H_k}\left(\mathbf{I} - \frac{\mathbf{\delta g} \mathbf{\delta x^T}}{\mathbf{\delta x^T} \mathbf{\delta g}} \right) + \frac{\mathbf{\delta x} \mathbf{\delta x^T}}{\mathbf{\delta x^T} \mathbf{\delta g}} \]
        \State \(\mathbf{x} \gets \mathbf{x_{\text{new}}}, \quad k \gets k+1\)
    \EndWhile
    \State $result.status = 0$
    \State \Return \(result\)
\EndFunction
\end{algorithmic}
\end{algorithm}

The BFGS \cite{broyden1970convergence,fletcher1970new,goldfarb1970family,shanno1970conditioning} algorithm is a well-known quasi-Newton method that approximates the inverse of the Hessian matrix using a combination of rank-1 updates shown in Algorithm~\ref{alg:sequential-bfgs}. 
Algorithm~\ref{alg:sequential-bfgs} also summarizes the exact update that is used to calculate $H_{k+1}$ at each iteration.
We calculate the step size $\alpha$ using a backtracking line search described in Algorithm~\ref{alg:line-search}.


Multiple optimizations increase the likelihood of converging to the global minimum.
After running the optimizations from many points, we aggregate the results at each iteration to get the best one. 


\subsection{Automatic Differentiation}
Our implementation of forward-mode AD relies on dual numbers to compute gradients accurately and efficiently. 
Dual numbers are written as
\[
a + b\,\epsilon,
\]
where \(a\) and \(b\) are real numbers and \(\epsilon\) is a symbol with the property
\[
\epsilon^2 = 0, \epsilon \neq 0.
\]
This property means that any term involving \(\epsilon^2\) vanishes. When using dual numbers for automatic differentiation, the coefficient \(b\) represents the partial derivative of a function.
Then, we evaluate the objective function using our overloaded operators for dual numbers. 
Algorithm~\ref{alg:forwardad} describes the implementation of this approach by setting the value to $x[i]$ and the tangent or derivative part to 1.

\begin{algorithm}[ht]
\caption{Forward-Mode AD Procedure} 
\begin{algorithmic}[1]
\Function{forwardAD}{$f,\, x \in \mathbb{R}^{\text{dim}}$}
    \State Initialize $\mathrm{gradient[dim]}$
    \For{$i\gets0$ to $\text{dim}-1$}
         \State $xDual[i] \gets x[i] + 0\,\epsilon$ \Comment{initializing dual number}
    \EndFor
    \For{$i\gets0$ to $\text{dim}-1$}
         \State Set $xDual[i].dual \gets 1$ \Comment{Seed derivative for variable $i$} 
         \State $result \gets f(xDual)$
         \State $\text{gradient}[i] \gets result.dual$
         \State Reset $xDual[i].dual \gets 0$
    \EndFor
    \State \Return $gradient$
\EndFunction
\end{algorithmic}
\label{alg:forwardad}
\end{algorithm}


\subsection{Line Search}
The main goal of the line search in optimization algorithms is to select an optimal step size, \(\alpha\), that minimizes the objective function along a given search direction \(\mathbf{p}\).
The choice of this algorithm greatly influences the outcome of the optimization. 
Our algorithm implements a commonly used backtracking line search with the Armijo condition \cite{armijo1966minimization}, where it initially starts with a step size of 1. 
For the current point $x^{(i)}$, we aim to find the \(\alpha\) such that the new point \(\mathbf{x^{(i)}}_{\text{new}} = \mathbf{x^{(i)}} + \alpha^{(i)}\, \mathbf{p^{(i)}}
\)
yields the greatest reduction in function value \(f(\mathbf{x^{(i)}})\).

The Armijo condition \cite{armijo1966minimization} for backtracking is:
\[
f(\mathbf{x} + \alpha\,\mathbf{p}) \leq f(\mathbf{x}) + c_1\,\alpha\, (\nabla f(\mathbf{x})^\top \mathbf{p}),
\]
where \(c_1\) is a small constant that we fixed to $0.3$. The line search we used relies on this balance between large steps for faster progress and little steps to avoid overshooting.
We found that twenty iterations is sufficient, and therefore we fixed that parameter. 
The following section gives an overview of the approach to GPU parallelization.

\begin{algorithm}[ht]
\caption{Backtracking line search using Armijo condition}
\begin{algorithmic}[1]
\Function{lineSearch}{$f,\mathbf{x},\mathbf{p},\mathbf{g}, \mathrm{iter}_{ls}$}
    \State \(c_1 \gets 0.3, \alpha \gets 1.0\)
    \State \(ddir \gets \Call{dotProduct}{\mathbf{g}, \mathbf{p}}\)
    \For{\(i \gets 0\) to $\mathrm{iter}_{ls}$}
        \State \(\mathbf{xTemp} \gets \mathbf{x} + \alpha \, \mathbf{p}\)
        \State \(f_1 \gets \text{$f$}(\mathbf{xTemp})\)
        \If{\(f_1 \leq f_0 + c_1 \, \alpha \, ddir\)}
            \State \textbf{break}
        \EndIf
        \State \(\alpha \gets 0.5 \times \alpha\)
    \EndFor
    \State \Return \(\alpha\)
\EndFunction
\end{algorithmic}
\label{alg:line-search}
\end{algorithm}

\section{Approach to GPU parallelization} \label{sec:four}
This section describes how we utilize parallelism to speed up each of the components of the sequential algorithm. 
In Algorithm~\ref{alg:sequentialzeus}, we parallelize the naturally independent phases of the sequential algorithm.
The PSO initialization and each iteration can be done in parallel, and then the BFGS for each independent optimization. 
In both Algorithm~\ref{alg:sequentialzeus}, and Algorithm~\ref{alg:zeus-pso} the loop starting in Line 6 cannot be parallelized since each iteration depends on its previous iterations' velocities and best locations. 
Each thread randomly initializes the particles, runs the PSO main loop, and then applies BFGS until enough optimizations have converged.
For this study, the parallelism we explored is on the first level. 
More levels of parallelism will be further studied in future work.

\begin{algorithm}[ht]
\begin{algorithmic}[1]
\caption{{\sc Zeus} Parallel PSO‐BFGS}
\label{alg:zeus-pso}
\Procedure{{\sc Zeus}}{$f,N,\mathrm{range},\mathrm{iter}_{bfgs},\mathrm{iter}_{pso},\mathrm{iter}_{ls},\allowbreak required_c,\Theta, w, c_1, c_2$}
    \State Allocate $\mathbf{swarm}[N*dim], \mathbf{V}[N*dim], \mathbf{pX}[N*dim], \mathbf{pB}[N], \mathbf{gX}[dim]$;
     \State $\mathrm{converged} \gets 0$;  \Comment{number of converged runs}
    \State $gF \gets +\infty$;
    \State  \Call{psoInitKernel}{$f,\mathbf{swarm},\mathrm{range},\mathbf{V},\mathbf{pX},pF, \mathbf{gX},\allowbreak gF,N$}
        \For{$i=0$ to $\mathrm{iter}_{pso}$}
            \State \Call{psoIterKernel}{$f,\mathbf{swarm},\mathrm{range},\;w,c_1,c_2,\allowbreak \mathbf{V},\mathbf{pX},pF,\mathbf{gX},gF,N$}
        \EndFor
    \State \Call{BFGSKernel}{$f, \mathrm{range},\mathrm{iter}_{bfgs},\mathrm{iter}_{pso},\mathrm{iter}_{ls},\mathbf{swarm},\allowbreak\Theta,\mathrm{required}_c, \mathrm{converged}, stopflag$}
    \State $(best,\;idx)\gets \Call{ParallelReduction}{\{\mathbf{swarm}[i]\}_{i=1}^N}$
    \State \Return $best$
\EndProcedure
\end{algorithmic}
\end{algorithm}

\subsection{Parallel RNG}
To initialize the starting points in Algorithm~\ref{alg:bfgs-kernel} we generated the random numbers on the GPU at runtime using the cuRAND library \cite{cook2012cuda} by NVIDIA. 
We integrated an on-demand strategy where each thread generates its own starting point at runtime. 
For each dimension, a thread produces one double-precision number uniformly sampled from a user-given range $[lower,upper]$ using a unique seed based on its thread's global index plus the current dimension.
This approach eliminates the need to store the entire size of $N \times dim$ array in memory.

\subsection{Parallel PSO}
The way we handled parallelism can be broken down into two main phases: (1) the initialization of the particles and (2) the iterations of the swarm. 
Algorithm~\ref{alg:gpu-pso-init-kernel} initializes the swarm in parallel for each particle and determines the global best using atomic operations. 
Algorithm~\ref{alg:gpu-pso-iter-kernel} is being used to update positions and velocities for each particle and find the global best at each iteration using atomic operations. 
Intuitively, it might seem that more iterations would yield superior accuracy.
In Section~\ref{sec:pso_iter} we show that this is not the case for all function types.

\begin{algorithm}[ht]
\begin{algorithmic}[1]
\caption{PSO Initialization Kernel}
\label{alg:gpu-pso-init-kernel}
\Procedure{psoInitKernel}{$f, \mathbf{swarm}, \mathrm{range}, \mathbf{V}, \mathbf{pX},pF, \mathbf{gX}, gF, N$}
    \For{$i \gets 0$ to $N-1$ \textbf{in parallel}}
        \If{$i \ge N$} \State \textbf{return} \EndIf
        \State $\Delta v \gets \,(\mathrm{upper}-\mathrm{lower})$
        \State $\mathbf{r_x} \gets \mathrm{rand}(s,\mathrm{lower},\mathrm{upper})$
        \State $\mathbf{r_v} \gets \mathrm{rand}(s,-\Delta v,\Delta v)$
        \State $\mathbf{swarm[i]}\gets r_x$; \quad $\mathbf{V[i]}\gets \mathbf{r_v}$; \quad $\mathbf{pX}\gets \mathbf{r_x}$
        \State $fval \gets f.\mathrm{evaluate}(\mathbf{swarm[i]})$ 
        \State $pF[i] \gets fval$
        \State $old \gets \mathrm{atomicMin}(gF,\,fval)$ \Comment{atomic global‐best}
        \If{$fval < old$}
            \State $\mathbf{gX} \gets \mathbf{pX[i]}$ 
        \EndIf
        \State $\mathrm{states}[i]\gets s$
    \EndFor
\EndProcedure
\end{algorithmic}
\end{algorithm}

\begin{algorithm}[ht]
\begin{algorithmic}[1]
\caption{PSO Iteration Kernel}
\label{alg:gpu-pso-iter-kernel}
\Function{psoIterKernel}{$f,\mathbf{swarm},\mathrm{range},\;w,c_1,c_2,\mathbf{V},\mathbf{pX},pF,\mathbf{gX},gF,N$}
    \For{$i \gets 0$ to $N-1$ \textbf{in parallel}}
        \If{$i \ge N$} \State \textbf{return} \EndIf
        \State $\mathbf{r_1} \gets \mathrm{rand}(s,0,1), \mathbf{r_2} \gets \mathrm{rand}(s,0,1)$
        \State $\mathbf{x} \gets \mathbf{swarm[i]}, \mathbf{v} \gets \mathbf{V[i]}$
        \State $\mathbf{p} \gets \mathbf{pX[i]},\;\mathbf{g} \gets \mathbf{gX}$
        \State $\mathbf{v'}\gets w\,\mathbf{v} + c_1\,\mathbf{r_1}\,(\mathbf{p}-\mathbf{x}) + c_2\,\mathbf{r_2}\,(\mathbf{g}-\mathbf{x})$
        \State $\mathbf{x'} \gets\mathbf{x} + \mathbf{v'}; \quad \mathbf{V[i]} \gets \mathbf{v'} \quad \mathbf{swarm[i]} \gets \mathbf{x'}$
        \State $fval \gets f.\mathrm{evaluate}(\mathbf{swarm[i]})$
        \If{$fval < pF[i]$}
            \State $pF[i]\gets fval$;\quad $\mathbf{pX[i]}\gets \mathbf{swarm[i]}$
        \EndIf
        \State $old \gets \mathrm{atomicMin}(gF,\,fval)$
        \If{$fval < old$}
            \State $\mathbf{gX} \gets \mathbf{swarm[i]}$
        \EndIf
        \State $\mathrm{states}[i]\gets s$
    \EndFor
\EndFunction
\end{algorithmic}
\end{algorithm}

\subsection{Parallel BFGS}
Our approach executes a single optimization per thread, utilizing the extremely parallel nature of GPUs. 
Each thread gets a piece of the entire swarm array of size $N \times dim$.
By starting from many points in parallel, our aim is to explore a large portion of the search space simultaneously. 

In this implementation, we are synchronizing the threads so that if we hit the number of required converged optimizations, we let all threads know after to prevent wasting resources.
At each iteration, we check to see if anyone has hit the $stopFlag$.
Once the stop flag is hit, each thread currently working will see it at the next iteration.
Then, we used CUDA's reduction kernel to find the global best from all of the results, as it provides an easily accessible kernel function\cite{cook2012cuda}.

We use the same line search and forward AD method as in the sequential algorithm.
While the AD stage can in principle be parallelized because each partial derivative is independent, our performance measurements show that the Hessian update step dominates the BFGS kernel runtime as the problem dimension increases.  
In contrast, the AD component accounts for only a small fraction of the total runtime. 



\begin{algorithm}[ht]
\begin{algorithmic}[1]
\caption{BFGS Kernel with forward-mode AD}
\label{alg:bfgs-kernel}
\Function{BFGSKernel}{$f, \mathrm{range},\mathrm{iter}_{bfgs},\mathrm{iter}_{pso},\mathrm{iter}_{ls},\allowbreak \mathbf{swarm},\Theta,required_c, \mathrm{converged}, stopflag$}
    \For{$i \gets 0$ to $N-1$ \textbf{in parallel}}
        \State \(\mathbf{x} \gets \mathbf{swarm}[i]\), \(k \gets 0\) \Comment{Set initial guess, loop counter}
        \State  \(\mathbf{H} \gets \mathbf{I}\) \Comment{Identity matrix for Hessian}
        \While{\(k < \text{max\_iter}\)}
            \If{$\text{atomicAdd}(stopFlag,0) \neq 0$}        
                \Comment{someone hit the stop flag}
                \State \textbf{break}
            \EndIf
            \State \(\nabla f(x) \gets \) \Call{forwardAD}{$f$, $\mathbf{x}$}
            \If{\(\|\nabla f(x)\| <\Theta\)}
                \State $\,\mathrm{old} \gets \text{atomicAdd}(converged,1)$
                \If{$\mathrm{old} = \mathrm{required_c}$}
                    \State $\text{atomicExch}(stopFlag,1)$      
                    \Comment{first thread to reach the target sets stop}
                \EndIf
                \State \textbf{break} \Comment{convergence criterion met}
            \EndIf
        \State \(\mathbf{p} \gets -\mathbf{H}\, \nabla f(\mathbf{x})\) \Comment{Calculate search direction}
        \State \(\alpha \gets \Call{lineSearch}{f(\mathbf{x}),\, \mathbf{x}, \mathbf{p}, \mathbf{g},\mathrm{dim},\mathrm{iter}_{ls}}\)
        \State \(\mathbf{x_{\text{new}}} \gets \mathbf{x} + \alpha\, \mathbf{p}\) \Comment{Update current point}
        \State \(\mathbf{\delta x} \gets \mathbf{x_{\text{new}}} - \mathbf{x}\) \Comment{Compute differences}
        \State \(\mathbf{\delta g} \gets \nabla f(\mathbf{x_{\text{new}}}) - \nabla f(\mathbf{x})\)
        \State Hessian update:
        \[\mathbf{H}_{k+1} \gets \left(\mathbf{I} - \frac{\mathbf{\delta x} \mathbf{\delta g^T}}{\mathbf{\delta x^T} \mathbf{\delta g}} \right) \mathbf{H_k}\left(\mathbf{I} - \frac{\mathbf{\delta g} \mathbf{\delta x^T}}{\mathbf{\delta x^T} \mathbf{\delta g}} \right) + \frac{\mathbf{\delta x} \mathbf{\delta x^T}}{\mathbf{\delta x^T} \mathbf{\delta g}} \]
        \State \(\mathbf{x} \gets \mathbf{x_{\text{new}}}, \quad k \gets k+1\)
    \EndWhile
        \State \Return \(x\)
    \EndFor
\EndFunction
\end{algorithmic}
\end{algorithm}

\section{Experiments and Results} \label{sec:five}
\subsection{Experimental Setup}
In this subsection we provide details about the experimental setup, including hardware and software used throughout the study. 
We used the same compute cluster to obtain both sequential and parallel results, where we used an Intel Xeon Gold 6148 CPU @ 2.40 GHz.
On this same server we allocated an NVIDIA A100 device with 80GB of VRAM. 
This device has 6912 CUDA cores. 
To compile our code, we used CUDA~12.1.

We have utilized four commonly used benchmark functions described in the next subsection.
For each function the error was calculated using the Euclidean distance that measures the distance between the estimated coordinates and the actual coordinates of the global minimum. \label{sec:euclidean}
When identifying a desirable error, we set the threshold to be $10^{-6}$. \label{sec:tolerance}
The performance of our algorithm depends mainly on the characteristics of the functions being optimized and the number of dimensions. 
For this reason, we experimented with four widely used objective functions to test the performance of baseline methods and our approach.

\subsection{Test Functions} \label{sec:testfns}
The following test functions try to cover the spectrum of test functions.
On convex surfaces like the Rosenbrock and Golstein-Price functions, our algorithm converges using a single optimization. 
However, for surfaces with many local minima for the Rastrigin and Ackley functions, many optimizations are needed to converge, especially in high-dimensional spaces.  

\subsubsection{Rosenbrock}
The Rosenbrock function was proposed in 1960 and has been widely used to test the effectiveness of mathematical optimization algorithms \cite{rosenbrock1960automatic}. 
It is a summation, where the number of dimensions can be increased dynamically, and is given by:
\[
f(\mathbf{x}) = \sum_{i=1}^{n} \left[ (1 - x_i)^2 + 100 \cdot (x_{i+1} - x_i^2)^2 \right]
\]
where $\mathbf{x} = (x_1, x_2, \dots, x_{n})$ and $\mathbf{x} \in \mathbb{R}^{n}$

\subsubsection{Rastrigin}
The Rastrigin test function was first proposed in 1974 \cite{rastrigin1974systems}. 
It is often used to test the efficiency of optimization algorithms in terms of convergence to the global minimum and execution time due to its highly multimodal nature. 
The number of local minima grows exponentially with the number of dimensions, which is shown in Figure~\ref{fig:multistart4solution}.

The Rastrigin function in $N$ dimensions is defined as:
\[
f(\mathbf{x}) = A \cdot N + \sum_{i=1}^{N} \left[ x_i^2 - A \cdot \cos(2\pi x_i) \right]
\]
$$\text{where } A = 10, \quad \mathbf{x} \in \mathbb{R}^N$$
The function has a periodic pattern with local minima occurring at integer coordinates.
The global minimum is at the origin. 
In the range, $[-5.12, 5.12]$ there are 11 integer values, which means in this space there are $11^{2}$ or $121$ local minima, with one being the global minimum. 

\subsubsection{Ackley}
The Ackley function is a widely used function to test mathematical optimization algorithms. 
The function was proposed by David Ackley \cite{ackley2012connectionist}. 
It is defined for general d-dimensions:
\[ 
f(x) = \begin{aligned} -20 \exp\Bigl(-0.2 \sqrt{\frac{1}{d}\sum_{i=1}^{d} x_i^2}\Bigr) - \exp\Bigl(\frac{1}{d}\sum_{i=1}^{d}\cos(2\pi x_i)\Bigr) \\ 
+ e + 20 
\end{aligned}
\]

It has a similar landscape to the Rastrigin function with many local minima due to the periodic nature of the cosine function.
In 2 dimensions, it has a single global minimum at (0.0, 0.0) where the function value is 0.
However, the derivatives are undefined at (0.0, 0.0).
For functions of this type we acknowledge the ``failure" mode of our algorithm, where the program returns diverged status, where we reach the maximum number of iterations without converging based on the \(\|\nabla f(x)\| <\Theta\) criterion.  
This behavior is illustrated in Figure~\ref{fig:ackley}.
Future work include resolving this issue.
This function violates the condition on continuity of derivatives. 
The algorithm never knows it has converged, never reaches a location where \(\|\nabla f(x)\| < \Theta\).


\subsubsection{Goldstein-Price}
This function was first developed by Goldstein \cite{goldstein1971descent}.
It has a single minimum, but many starts require an infeasible number of steps to converge.
It is defined as:
\[
\begin{aligned}
f(\mathbf{x}) =
    \bigl[1 + (x_1 + x_2 + 1)^2\,(19 - 14x_1 + 3x_1^2 - 14x_2 \\ + 6x_1x_2 + 3x_2^2)\bigr]
&\quad \times \\
    \bigl[\,30 + (2x_1 - 3x_2)^2 (18 - 32x_1 + 12x_1^2 + 48x_2 \\ - 36x_1x_2 + 27x_2^2)\bigr]
\end{aligned}
\]

Using these test functions, we have conducted experiments of two flavors. 
The first one describes the effectiveness of the algorithm, which is the ability to find the global minimum, the use of PSO, and BFGS. 
The second describes the GPU performance benefits. 
There is a need for multistart for several reasons. 

\subsection{Finding the right solution} 

\begin{figure}
    \centering
    \includegraphics[width=\linewidth]{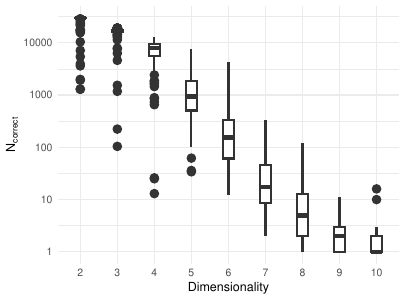}
    \caption{Box and whisker plot showing performance degrades drastically for the Rastrigin function as the dimensionality of the problem increases when using the same number of particles. For each dimension, we plot 100 runs, where each run is a result of using $10^5$ particles and 5 PSO iterations. We count the number of correct solutions across each run. $N_{\mathrm{correct}}$ corresponds to the count of each optimizations where the Euclidean error is less than 0.5.}
    \label{fig:multistart4solution}
\end{figure}


In this experiment, we demonstrate the need for multistart approaches for functions with many local minima. 
If there are multiple local minima in the hyperspace like the Rastrigin function, as dimensionality grows we need more and more starting points to be confident in convergence to the global minimum.
The 5-dimensional variant has $11^5$ or $161,051$ local minima with a single global minimum. 
For the 10-dimensional Rastrigin function, the number of local minima in our searching space is $11^{10}$ or 26 billion, which means we would require to launch vastly more particles. 
A practitioner using the {\sc Zeus} algorithm would also need more than one convergences to have confidence we have the correct one.
We describe future work how to handle this. 

In Figure~\ref{fig:multistart4solution}, we plot the distribution of a 100 runs using box and whiskers for each dimension, the number of particles that converge into the basin of the true global minimum.
The plot shows results where each run went until the algorithm claims convergence 100 times in different dimensions of the Rastrigin function. 
The distribution of successful counts rapidly decreases with each added dimension, showing the exponential growth of local minima. 
For the first couple dimensions, the box is a line, as the distribution of the 100 runs are close together.
With each dimension, a practitioner should trust the solution less and less. 
By ten dimensions, the number of correct solutions is effectively zero.



\begin{figure}
    \centering
    \includegraphics[width=0.99\linewidth]{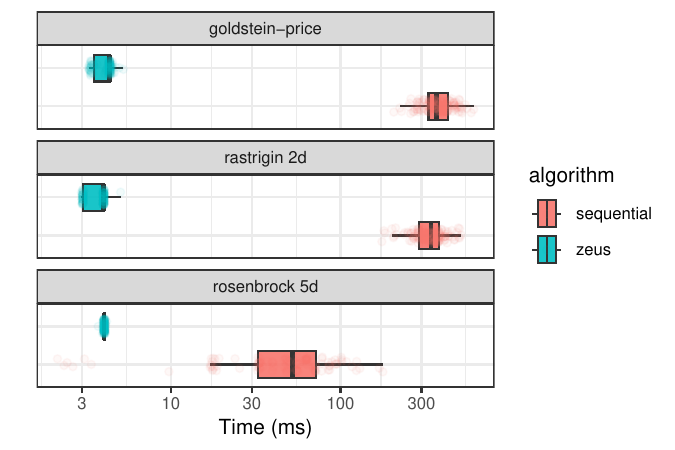}
    \caption{Visual illustration of the speed advantage achieved by {\sc Zeus} for 2-dimensional and 5-dimensional objective functions. CPU runtimes were divided by the number of cores to approximate the ideal parallel execution. The distributions are based on 100 runs. Vertical jitter was applied to each point to make them more visible for {\sc Zeus}. The Ackley function was left out due to its misbehavior shown in Figure~\ref{fig:ackley}.}
    \label{fig:speedup}
\end{figure}

\subsection{Parallel multistart for speed} 
By running many starts in parallel and stopping once a set number have converged, {\sc Zeus} cuts the time by orders of magnitude versus the fully sequential implementation.
Running the sequential algorithm becomes infeasible for anything greater than a 2-dimensional Rastrigin function because it requires too many starts. 
For functions like the 2-dimensional Rosenbrock and Goldstein-Price function, BFGS will eventually converge from anywhere if we let it run long enough. 

One of the advantages of the parallel algorithm is that it does not suffer from individual starting points that are far from the right solution.
Whereas in the sequential variant, we must keep looping until we have enough converged optimizations. 
As a result, we can observe multiple orders of magnitude difference in the time it takes to converge for 100 optimizations using a small swarm of 1024.
Figure~\ref{fig:speedup} visually demonstrates this speedup between the sequential and parallel {\sc Zeus}.
For fair comparison, the times for the sequential algorithm were divided by the number of cores of the CPU used for this experiment. 
The box and whisker plot shows a large difference for the 2-dimensional problems that only increases with the dimensionality of the problem. 
However, for the {\sc Zeus} algorithm, it is wasteful to use 1024 threads, as we are leaving many cores idle.

\begin{figure}
    \centering
    \includegraphics[width=\linewidth]{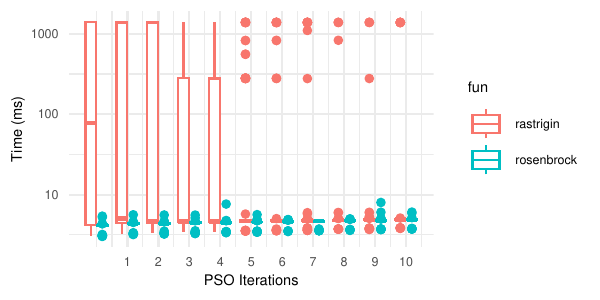}
    \includegraphics[width=\linewidth]{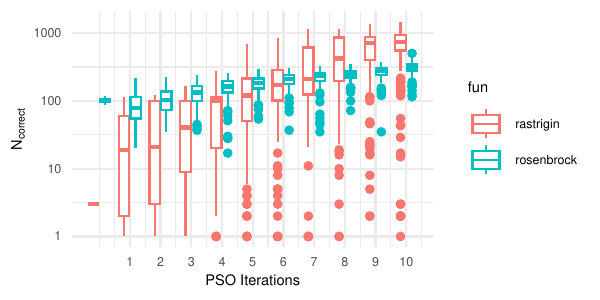}
    \caption{Performance plots in terms of time (top) and number of correct solutions (bottom) across 100 runs compared with the PSO iterations for 5-dimensional Rastrigin (orange) and Rosenbrock (blue) functions. The Rastrigin function in this dimension has $11^{5}$ or $161,051$ local minima. $N_{\mathrm{correct}}$ corresponding to the count of each optimization where the Euclidean error is less than 0.5.}
    \label{fig:pso_iter}
\end{figure}

\subsection{Improving the starting points by PSO} \label{sec:pso_iter}
In this experiment, we show that it is useful to do a handful of PSO iterations as we increase the number of correct solutions and decrease the time it takes to converge to the required number of convergences. 
We show which functions benefit more from PSO iterations. 
Once we have enough converged particles where the norm of the gradient is sufficiently small enough, then BFGS synchronizes all other threads to stop early. 
Figure~\ref{fig:pso_iter} demonstrates that PSO helps optimize functions with many local minima, and it is not too much of a waste of resources for problems with landscapes that have flat regions. 
In practical use cases, the user does not have information about the landscape of the function being optimized. 
The plot illustrates that for the Rastrigin function, the time it takes to gather enough convergences goes down with the number of PSO iterations (top), whereas the Rosenbrock function still benefits from more PSO computation in terms of the number of correct solutions (bottom).
We can observe that we find more correct solutions as we increase the number of PSO iterations. 
For the Rastrigin function, we increased the number of correct solutions by multiple orders of magnitude. 




\subsection{Comparison with other libraries}
\begin{figure}
    \centering
    \includegraphics[width=\linewidth]{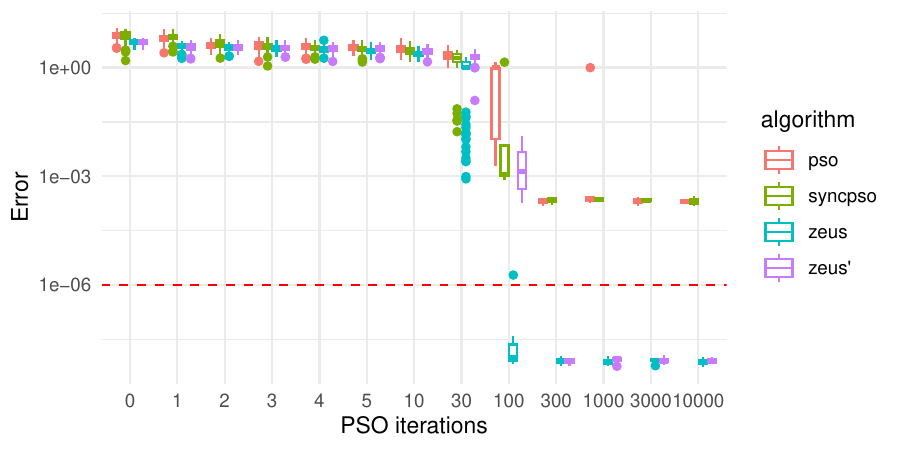}
    \includegraphics[width=\linewidth]{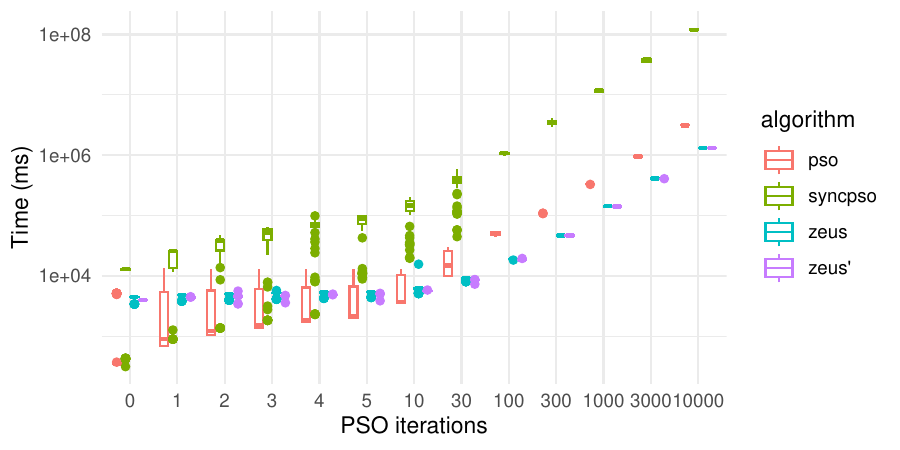}
    \caption{Comparative performances for the 10-dimensional Rastrigin function with $11^{10}$ local minima. For a 10-dimensional Rastrigin to have 1 for the Euclidean error, means the point landed in a local minima, not the local minima.}
    \label{fig:rast10}
\end{figure}

The section compares the algorithm's performance against a Julia library in terms of error and time, and discusses how multi-start algorithms can utilize parallel execution on GPUs for faster convergence.
In order to make the problem more complex, we have increased the number of dimensions to be 10 for this experiment.
With this experiment we aim to gain insight into convergence behavior as we increase the number PSO iterations. 
We were able to use two different algorithms from their library. 
One of them is a parallel PSO with a synchronization point at each step, the other is a method that they note has race conditions. 
We do not consider using the asynchronous variant because it lacks the global update
We can observe from Figure~\ref{fig:rast10} that the algorithm has a degraded accuracy as we increase the number of steps compared to its other variant.
This might be related to the race condition they mention, but further investigation should be done to confirm.
After extensive communication with the authors, we were unable to get their Hybrid algorithm that uses BFGS to function.
Since they have hard-coded the hyperparameters, we use the same ones to draw a fair comparison.
We mark this algorithm as {\sc Zeus'}.
The {\sc Zeus} algorithm uses hyperparameters derived from a paper \cite{deboucha2020modified}.

From the figure we can observe that {\sc Zeus} outperforms the library that has two of the same ingredients. 




\subsection{Real-world application}
\begin{figure}
    \centering
    \includegraphics[width=\linewidth]{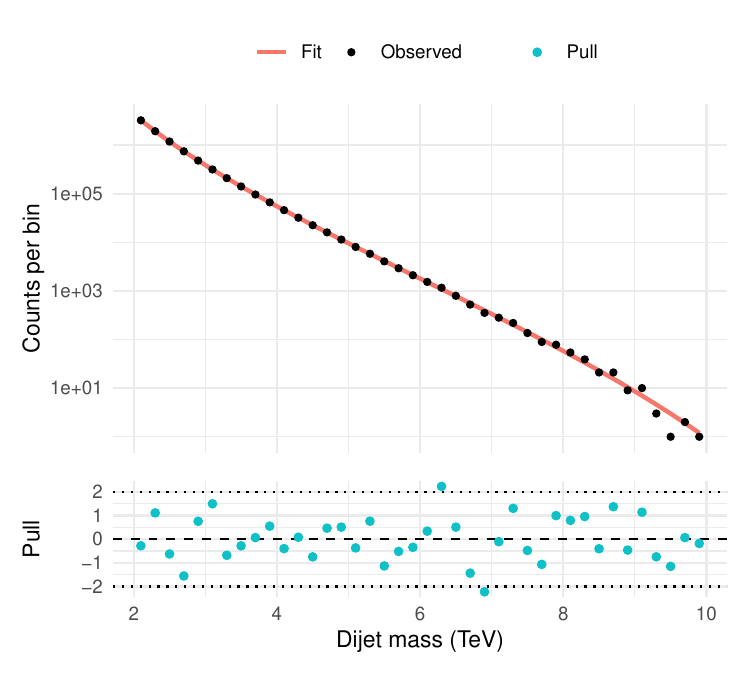}
    \caption{Simulated dijet mass spectrum and fitted dijet mass spectra.
    The top panel shows the simulated event counts (black points) compared with the fitted prediction (red line). 
    The bottom panel plots the pull distribution (blue points), defined as $\frac{N_{\text{obs}} - N_{\text{pred}}}{\sigma}$, where $N_{\text{obs}}$ is the simulated count of events, $N_{\text{pred}}$ is the model prediction, and $\sigma$ is the statistical uncertainty per bin. 
    The pulls fluctuate around zero and lie mostly within $\pm2\sigma$, indicating  agreement between simulation and prediction.}
    \label{fig:dijet}
\end{figure}

An important real-world application of numerical optimization is model fitting: given a numerical model with a set of parameters, to determine the best set of parameters to fit the data.
The goal of such fitting is to constrain the model so that it can accurately predict future data.
One way to evaluate the quality of the fit is to compare the observed data with the fitted model's prediction.
In figure~\ref{fig:dijet} shows the fitting of a simulated dijet mass spectrum and evaluation of the quality of the fit. 
The top panel compares the observed count of events to the prediction, showing that {\sc Zeus} finds fit parameters that yield an accurate prediction of the simulated data.
The bottom panel shows the so-called \emph{pull} distribution, which quantifies the deviations of each bin from the prediction relative to its statistical uncertainty. 
The pulls are centered around zero, with most values within $\pm2\sigma$, indicating that the residuals are consistent with random statistical fluctuations and confirming the excellent quality of the fit.

\section{Problem with convergence} \label{sec:six}
\begin{figure}
    \centering
    \includegraphics[width=0.99\linewidth]{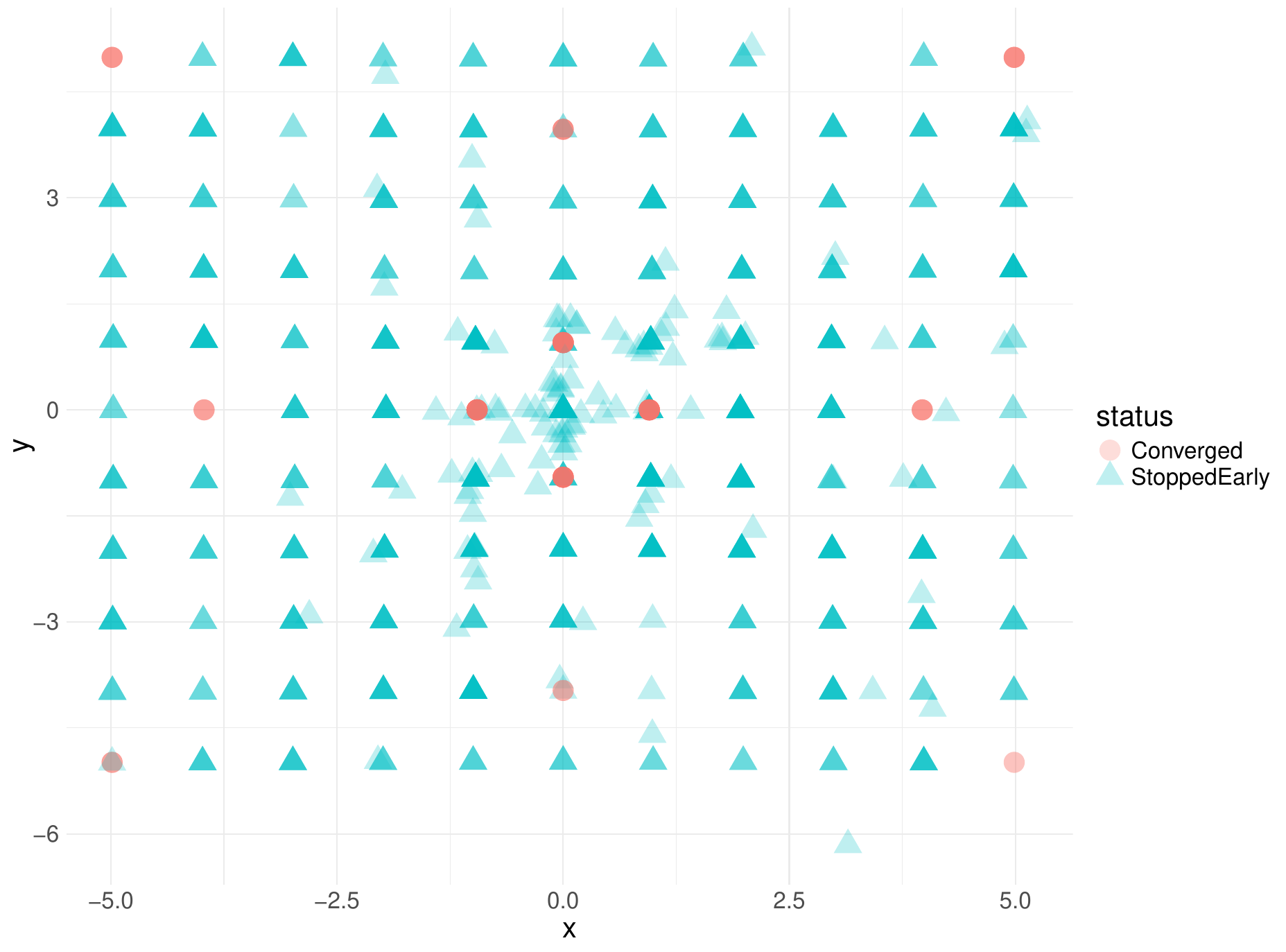}
    \caption{Misbehavior of the algorithm claiming convergence for a 2-dimensional Ackley function for points where the norm of the gradient is sufficiently small enough and not declaring convergence for particles that are near the global minimum.}
    \label{fig:ackley}
\end{figure}

BFGS is controlled by the criterion of \(\|\nabla f(x)\| <\Theta\). 
This condition works well for well-behaved functions with continuous first-order derivatives. 
For a function with discontinuous derivatives, the minimum may be located at a discontinuity.
If the user sets the threshold to be too small, our program returns that we have not found the global minimum by the condition of the norm of the gradient. 
This behavior is illustrated in Figure~\ref{fig:ackley}, where the algorithm claimed convergence for optimizations landing in local minima where the condition \(\|\nabla f(x)\| <\Theta\) was satisfied.
However, points in or near the basin of the global minimum are stopped early because their norm was not below the threshold set by the user.
Future work will handle functions with discontinuous derivatives. 

\section{Discussion} \label{sec:seven}

\subsection{Performance Analysis}
Our experiments demonstrated several clear patterns in how {\sc Zeus} behaves across different test functions. 
On the Rastrigin problem, we saw that as we increase the number of dimensions from two up to ten, the number of particles that actually find the global minimum falls close to zero. 
This happens because the number of local minima increases exponentially, which means the chance of landing in the correct basin rapidly vanishes. 

By running many BFGS optimizations in parallel and stopping as soon as a set number of threads have converged, {\sc Zeus} achieves significant speedups over the fully sequential approach.
In our experiment, the time dropped by one to two orders of magnitude on both two- and five-dimensional functions.
The higher the dimension, the more benefit we gain by the parallel algorithm.


We also observed that our simple gradient-norm based convergence criterion can misfire on functions with discontinuous derivatives like the Ackley function. 
In such cases, particles may declare convergence too early when the norm of the gradient is small enough, but the point still may be far from the true minimum. 
Addressing this will require more sophisticated stopping criteria to handle functions with discontinuous derivatives in future versions of {\sc Zeus}.

\subsection{Future Work}
Future work will focus on exploiting the additional parallelism available in the single BFGS computation and reducing the computational complexity of the BFGS kernel itself.
While parallelizing AD remains an option, our measurements show that the Hessian update dominates the runtime as the dimension grows.
One promising approach is to explore L-BFGS \cite{liu1989limited}, which reduces the complexity of the update step while maintaining quasi-Newton convergence behavior.
This could significantly improve scalability, though potentially at some cost in solution accuracy due to the limited curvature information.

We will also explore means of improving the information given to the user about the degree of certainty in the reliability of the solution.
By clustering candidate solutions found by the multistart algorithm, we can identify candidate regions for the local minima.
If enough particles have been found to converge to the candidate region that has the lowest function value, and if no lower function value has been found, then we can have greater confidence that the region in question is the global minimum.
We will consider using an iterative process that continues until a user-settable number of particles have converged to the same lowest region.
We will explore two methods of clustering of the solutions based on their function value or their coordinates.


\section{Conclusion}
We have developed a GPU-accelerated algorithm, {\sc Zeus}, that has two main phases. 
The first phase selects random starting points which are then improved by using a PSO algorithm to move the particles to more promising regions.
The second phase is a gradient-based BFGS algorithm that uses forward-mode AD to calculate the gradient at each iteration.
No algorithm is universally best for every function, and the effectiveness of our algorithm depends on the specific function.
When using a gradient-descent algorithm like BFGS, for problems in which the presence of multiple local minimal is suspected, the use of multiple starting points is required.
Increasing the number of local minima increases the number of starting points needed to achieve confidence that the global minimum has been found.
Random selection of starting points is typically used to provide a good probability for several of the BFGS searches to converge to the global minimum.
Our results show that a few iterations of the PSO can improve the random starting points leading to faster convergence of the BFGS algorithm.
For highly multimodal objective functions like the Rastrigin function, PSO increases the fraction of starts that land in the basin of starting points that lead to the global minimum, whereas for unimodal objectives like the Rosenbrock function, BFGS alone would be sufficient if we let it run long enough.
Our open-source implementation makes these strategies available on GPUs, allowing practitioners to minimize non-convex problems more quickly and confidently than single-threaded solvers.
\bibliographystyle{ACM-Reference-Format}
\bibliography{references.bib}

\end{document}